\documentclass[prb,showpacs,twocolumn]{revtex4}

\usepackage{graphicx}
\usepackage{dcolumn}
\usepackage{bm}

\begin{document}

\title{Variational QMC study of a Hydrogen atom in jellium

with comparison to LSDA and LSDA-SIC solutions}

\date{\today}

\author{Andrew I. Duff}
 \email{A.I.Duff@bris.ac.uk}
\author{James F. Annett}
 \email{james.annett@bristol.ac.uk}
\affiliation{H. H. Wills Physics Laboratory, University of Bristol, Bristol
  BS8 1TL, United Kingdom}
\affiliation{Daresbury Laboratory, Daresbury, Warrington, WA4 4AD,
United Kingdom}

\begin{abstract}

A Hydrogen atom immersed in a finite jellium sphere is solved using
variational quantum Monte Carlo (VQMC). 
The same system is also solved using density functional theory (DFT), 
in both the local spin density (LSDA) and self-interaction 
correction (SIC) approximations. The immersion energies calculated using these
methods, as functions of the background density of the jellium, 
are found to lie within $1eV$ of each other
with minima in approximately the same positions.
The DFT results show overbinding relative to the VQMC result.
The immersion energies
also suggest an improved performance of the SIC over the LSDA
relative to the VQMC results.
The atom-induced density is also calculated
and shows a difference between the methods,
with a more extended Friedel oscillation in the case of the VQMC result.

\end{abstract}

\pacs{02.70.Ss, 71.10.Ca, 71.15.Mb, 31.15.Ew}

\maketitle

\bibliographystyle{unsrt}

\section{Introduction}

The model system of an atom embedded in infinite jellium is
of interest because it can be used to describe the bonding of
an atom within a solid. The idea is that the
atom is embedded in a homogeneous electron gas which is
regarded as the sum of the density tails from all other atoms in the
solid spatially averaged over the unit cell of the atom in question. 
The immersion energy of the atom
(the total energy of the atom in jellium minus the energies of the
separate atom and jellium)
can then be regarded as a first order approximation to the cohesive energy
per atom of the solid. In fact, within the framework of the 
effective medium theory (EMT),\cite{EMT,Many-Atom}
one includes an additional term in this cohesive energy
which describes the electrostatic attraction between the density tails within
the unit cell and the Hartree potential set up by the atom.
This allows one to calculate cohesive properties of solids across
the periodic table and indeed one finds that the
Wigner Seitz radii, bulk moduli and cohesive energies follow the same
trends as the experimental values.

In this paper
the atom in jellium system is solved using the
variational quantum Monte Carlo (VQMC) method 
\cite{var_dif_review,cep_chester} and the 
local spin density (LSDA) 
\cite{Hohen-Kohn,Kohn-Sham,DFT_book,LSDA1,LSDA2} and
self-interaction correction (SIC) \cite{SIC} approximations of
density functional theory (DFT).
The particular system used is 
a Hydrogen atom immersed in jellium.
In order to solve this system using VQMC, the system must be of finite size
and therefore we approximate the atom in infinite jellium
with an atom in a finite jellium sphere centered on the atom. 
The quantities to be calculated include the 
total energies of both the Hydrogen in jellium
and the jellium sphere by itself, the
immersion energy and the atom-induced density (the density
of the Hydrogen atom in jellium minus the density of the jellium).
These quantities are to be calculated as a function of the positive
background density of the jellium, $n_0$, for a jellium sphere with a
fixed number of electrons.

VQMC calculations have already been attempted on
the system of a Hydrogen atom in jellium by Sugiyama {\it et al}.
\cite{Sugiyama} These calculations however
resulted in immersion energies which differed significantly from the
LDA results. Atom-induced densities calculated by Sugiyama {\it et al}.
were also found to differ significantly between these methods.
Our results are based upon a careful study of size effects
and show a much closer agreement between LSDA and QMC.

The system of a jellium sphere with no embedded atom has been solved using
QMC (in both the variational and diffusion variants) by Sottile {\it et al}.
\cite{sottile} Jellium spheres with up to 106 electrons were reported
and good agreement was found between the LSDA and the diffusion QMC (DQMC)
results.

Work on using a Hydrogen atom in a finite jellium sphere has also
been reported \cite{Hintermann} within the LSDA. It was shown that by
controlling the size of the jellium sphere, a good approximation to
Hydrogen in infinite jellium can be obtained.

SIC calculations have already been reported for the 
atom in infinite jellium system.\cite{puska2}
The EMT was applied, and
for atoms up to and including the $2p$ elements
the SIC cohesive energy was found to be higher than that of the LSDA.
The interpretation placed on this was that the SIC 
was correcting for the overbinding present in the LSDA.
Here we compare SIC with both LSDA and VQMC.

In the following section we first investigate the LSDA solution of a
Hydrogen atom in a finite jellium sphere. We examine the effect of changing
the jellium sphere size and establish choices of sphere which yield good
approximations to the infinite jellium solution. We then consider
in Sections \ref{sec:VQMC} and \ref{sec:SIC} the VQMC and SIC methods
for the same finite sphere system. Section \ref{sec:results} presents
a comparison of our VQMC results with those of the LSDA and SIC.

\section{\label{sec:LSDA}LSDA solution}

\subsection{Minimizing the energy functional}

We first solve the spin-polarized Kohn-Sham (KS) equations 
within the LSDA \footnote{For the main results we use the parameterization
by Perdew and Zunger.\cite{perdew-zunger} In the test calculations of
finite size effects in
Fig.~\ref{fig:hyd in fin jell imm en vs density rs 3.25}-~\ref{fig:fin_jell_density_profiles}
we used the Gunnarsson-Lundqvist
parameterization.\cite{gunn-lund}} for
an external potential set up by a Hydrogen ion and a sphere of positive
charge centered on the ion with charge density $n_0$ and radius $R_{jell}$. 
The system is overall charge neutral.

We spherically symmetrize the spin densities, $n^\sigma({\bf r})$, 
after each iteration of
the self-consistency cycle before recalculating the KS potentials.
The KS potential will therefore also be spherically symmetric and so
the KS equations
reduce to the radial Schr\"odinger equations.
The self-consistent solutions 
\footnote{We use the Broyden method of D. Johnson \cite{Broyden}
as implemented by Hettler {\it et al}. \cite{Broyden2}}
to these
equations minimize the LSDA energy functional

\[
E[n^{\uparrow},n^{\downarrow}]= \sum_{n,l,m,\sigma}\int  
{\psi_{nlm}^{\sigma *}}
({\bf r})(-\frac{1}{2} \nabla^2) {\psi_{nlm}^{\sigma}} ({\bf r})d{\bf r}+
\]
\[
\frac{1}{2} \int \int  
{n(r)n(r')
 \over {\vert {\bf r} - {\bf r}' \vert}} d{\bf r} d{\bf r}'+
\int v_{ext}(r)
n(r) d{\bf r}+
\]
\[ 
E_{xc}[n^{\uparrow},n^{\downarrow}]
+Z \int_{r=0}^{r=R_{jell}} { n_0 \over { r }} d{\bf r}
+\frac{3}{5} {N^{5/3} \over {r_s}}
\] 

\noindent where $\psi_{nlm}^{\sigma}({\bf r})$ are the
self-consistent KS orbitals, $n^\sigma(r)$ are the 
spherically symmetrized spin densities, $n(r)=\sum_\sigma
n^\sigma(r)$ is the total electron density and
$v_{ext}(r)$ is the external potential

\[ 
v_{ext}(r)=-{1 \over r} - \int_{r'=0}^{R_{jell}} {n_0 \over {\vert
{\bf r} - {\bf r}' \vert}} d{\bf r}'
\] 

The final two terms in the energy functional 
include the Coulomb repulsion between the
ion and the positive background and the self-Coulomb
repulsion of the positive background.
The electron density parameter, $r_s$ is
defined by $\frac{4}{3} \pi r_s^3 n_0 = 1$ and
$N=\frac{4}{3} \pi R_{jell}^3 n_0$ 
is the number of electrons in the jellium sphere.

The kinetic energy can be calculated either by numerically evaluating
the radial derivatives of the orbitals, or alternatively by
rearranging the KS equations and
substituting in for the Laplacian. In our calculations both methods 
are used and are found to give the same results.

The energy functional is also used to calculate the energy of the
separate jellium sphere and Hydrogen atom. The immersion energy is then
obtained by subtracting
these quantities from the Hydrogen atom in jellium energy.

\subsection{Jellium sphere sizes}

\begin{figure}
\centerline{{\scalebox{0.8}{\rotatebox{-90}{\includegraphics{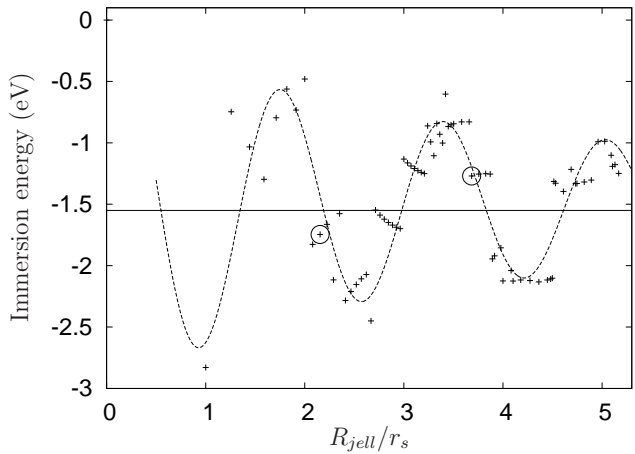}}}}}
\caption{\label{fig:hyd in fin jell imm en vs density rs 3.25}
The immersion energy versus jellium sphere radius for jellium spheres of 
background density $0.007a_B^{-3}$.
The straight line is the value of the immersion energy for Hydrogen
in infinite jellium of the same background density and
the sinusoidal line is a guide to the eye. The highlighted points
correspond to $N=10$ and $N=50$.
}
\end{figure}

The number of electrons in the sphere is chosen by studying,
within the LSDA, the
immersion energy of a Hydrogen atom in a jellium sphere of fixed $n_0$
as a function of the number of electrons in the sphere.
Figure~\ref{fig:hyd in fin jell imm en vs density rs 3.25} 
illustrates the dependence of the
immersion energy on the size of the jellium sphere for a background
density of $0.007a_B^{-3}$ ($r_s=3.25$).
The most obvious feature of the plot is the small scale
bunching of immersion energies, with the immersion energy increasing
or decreasing slightly
as a particular angular momentum shell is filled.
A larger scale feature is that the immersion energy oscillates around the 
immersion energy of Hydrogen in infinite jellium.\cite{Duff_thesis} 
These are Friedel oscillations and the wavelength predicted
by the theory is ${\Delta R / {r_s}} = {\pi / {(r_s k_F)}}
= {\pi / {(r_s (3 \pi^2 n_0)^{1 \over 3})}}  = 1.637$
which is in good agreement with the wavelength as read off
from Fig.~\ref{fig:hyd in fin jell imm en vs density rs 3.25}.

The amplitude of the Friedel oscillations in 
Fig.~\ref{fig:hyd in fin jell imm en vs density rs 3.25}
becomes smaller as the 
size of the jellium sphere is
increased. This tells us that we can make the immersion energy
arbitrarily close to the infinite jellium immersion energy by making
the jellium sphere very large.
However the damping of the oscillations is slow (unlike calculations by
Hintermann {\it et al}. \cite{Hintermann}).
Therefore because we are limited in the number of electrons we can include
in the VQMC calculation we must exploit the cross-over points at which
the immersion energy is as close as possible to the infinite jellium case.

We observe that
plots of immersion energy against $R_{jell}/r_s$, such as 
Fig.~\ref{fig:hyd in fin jell imm en vs density rs 3.25},
look the same for
different choices of $n_0$. In particular the same sinusoidal oscillation
is observed and the crossing points occur at the same values of 
$R_{jell}/r_s = N^{\frac{1}{3}}$.
Therefore if for a given choice of 
$n_0$, one chooses a value of $N$ for which the
immersion energy is close to that of the infinite jellium, then
one is assured that the immersion energy for all other values of $n_0$
will also be close to the infinite jellium immersion energy.

\begin{figure}
\scalebox{0.76}{\rotatebox{-90}{\includegraphics{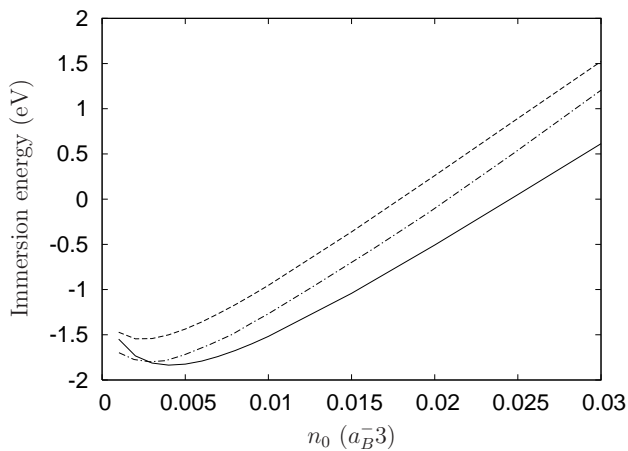}}}
\caption{\label{fig:hyd in fin jell imm en vs density}
Plots of immersion energy versus background density for a Hydrogen
atom immersed in jellium
spheres of size 10 (solid curve) and 50 (dashed curve) electrons. 
Also plotted is the immersion
energy curve for a Hydrogen atom in infinite jellium (dot-dash curve).}
\end{figure}

This is demonstrated by picking two values of $N$ which, from 
Fig.~\ref{fig:hyd in fin jell imm en vs density rs 3.25},
have immersion energies close
to the infinite jellium immersion energy. 
These values are $10$ and $50$, which are highlighted in 
Fig.~\ref{fig:hyd in fin jell imm en vs density rs 3.25}.
In Fig.~\ref{fig:hyd in fin jell imm en vs density} the immersion
energy is plotted for these two choices of $N$ for a range of $n_0$ values.
The curve for the 10-electron jellium sphere is seen to give a reasonable
approximation of the infinite jellium curve.\cite{Duff_thesis} 
However the 50-electron curve
improves further by giving the minimum in the same place
as for the infinite jellium. One sees that this curve is rigidly shifted
above that of the infinite jellium, confirming that the error 
in the immersion energy due to the finite size is approximately independent
of $n_0$.

\begin{figure}[t!]
\centerline{\scalebox{0.8}{\includegraphics{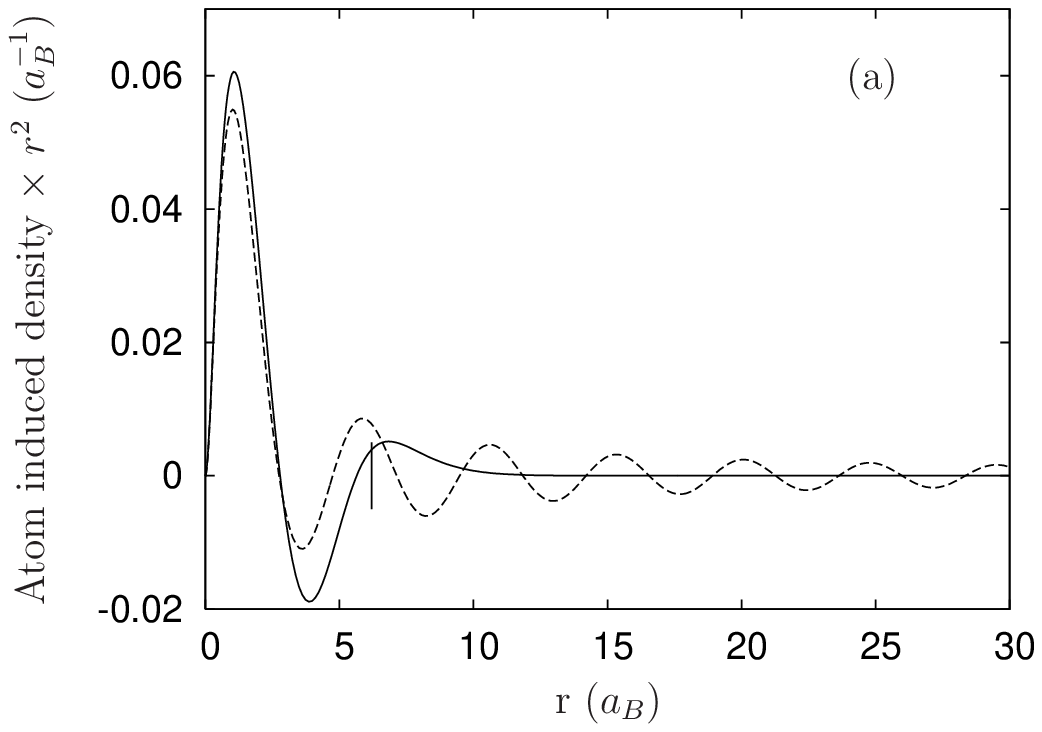}}}
\centerline{\scalebox{0.8}{\includegraphics{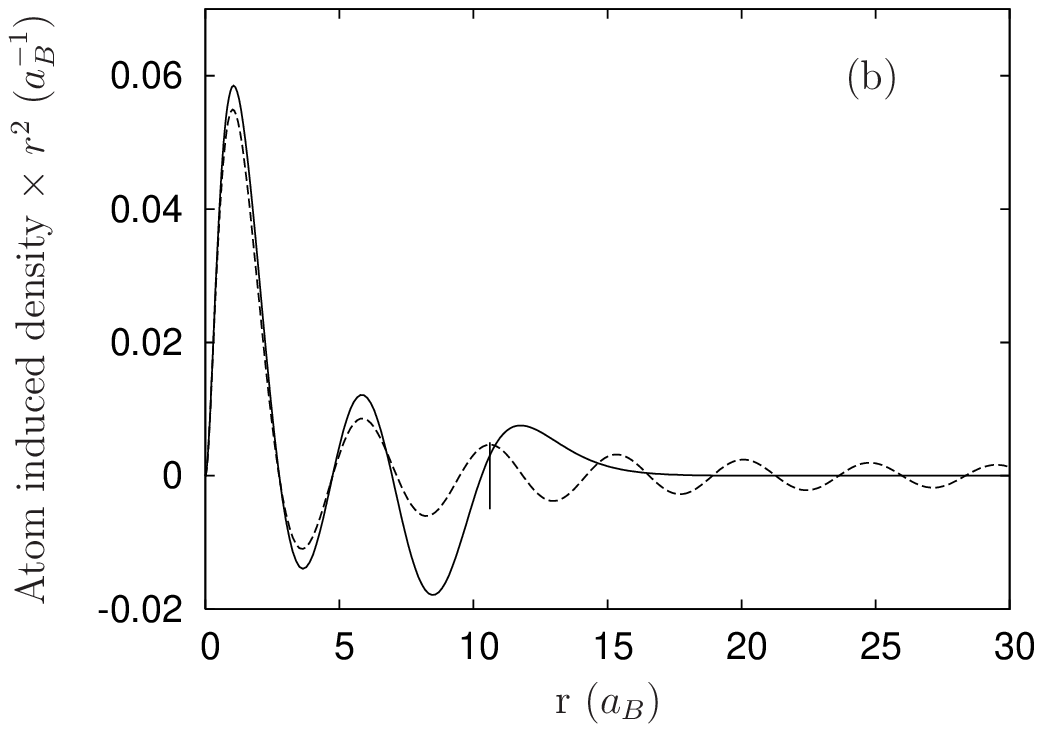}}}
\centerline{\scalebox{0.8}{\includegraphics{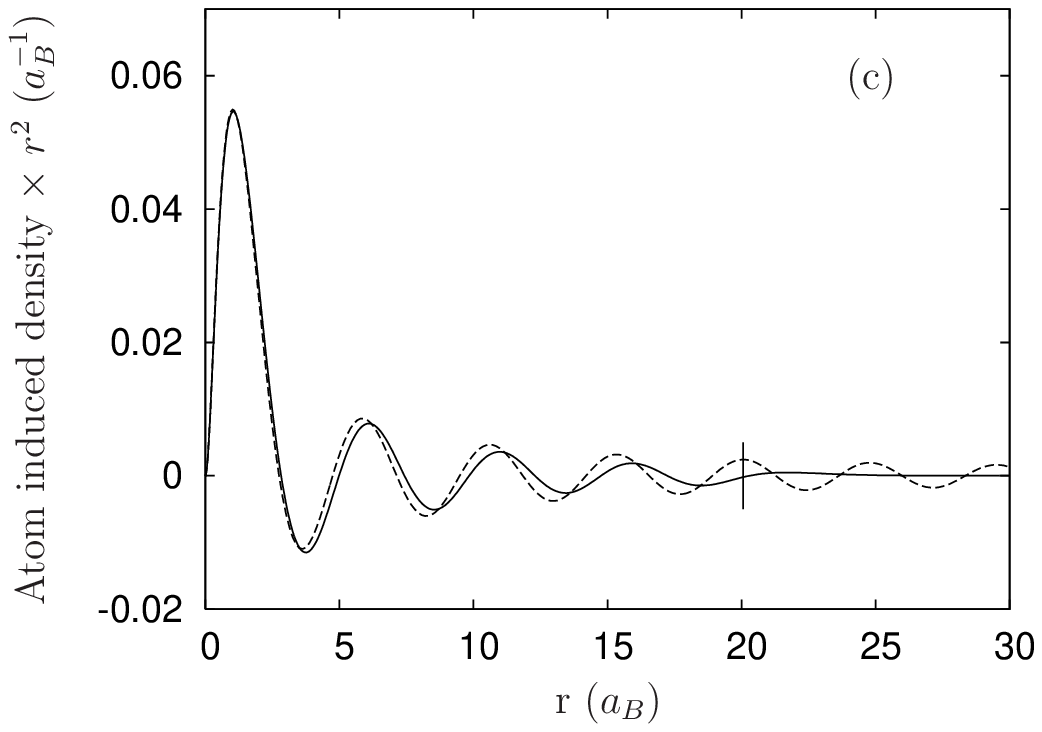}}}
\caption{\label{fig:fin_jell_density_profiles}
Plots of LSDA atom-induced densities for Hydrogen in finite jellium
spheres of background density $0.01a_B^{-3}$,
with (a) 10, (b) 50 and (c) 338 electrons (solid lines). 
Also plotted is the atom-induced
density for a Hydrogen atom in infinite jellium at the same
background density (dashed lines). The jellium sphere radius is
shown by a vertical straight line.}
\end{figure}

Our calculations also show
that the atom-induced density of the Hydrogen atom in finite
jellium tends smoothly 
towards that of a Hydrogen atom in infinite jellium as the size of the
jellium sphere is increased.
Figure~\ref{fig:fin_jell_density_profiles} shows our calculations of the atom-induced 
densities for a Hydrogen atom in jellium spheres
with a background density $0.01a_B^{-3}$. We consider
jellium spheres with $10$, $50$ and $338$ electrons and 
also plot the atom-induced 
density for a Hydrogen atom in infinite jellium. We see
clearly that as we increase the number of electrons in the jellium
sphere, the atom-induced density approaches the limiting atom-induced
density of a Hydrogen atom in infinite jellium.

\section{\label{sec:VQMC}VQMC Solution}

\subsection{Methodology}

In VQMC we use importance sampling to replace the
expectation value of the Hamiltonian with a sum over
so-called local energies

\[
\langle \Psi_T \vert \hat{H} \vert \Psi_T \rangle={1 \over {{\cal N}}} 
\int\Psi_T^*({\bf R})\hat{H}\Psi_T({\bf R})d{\bf R}=
\]
\[ 
\int {\vert \Psi_T({\bf R}) \vert^2
\over { {\cal N}}}
{\hat{H}\Psi_T({\bf R})
\over {\Psi_T({\bf R})}}d{\bf R} \approx
{1 \over {n}} \sum_{i=1}^n {\hat{H}\Psi_T({\bf R}_i)
\over {\Psi_T({\bf R}_i)}}
=\bar{E}
\] 

\noindent where ${\cal N}=\int \vert \Psi_T({\bf R}) \vert^2 d{\bf R}$,
${\bf R}_i$ are sets of particle coordinates, $\{ {\bf r}_1^{(i)},
{\bf r}_2^{(i)}, \dots, {\bf r}_N^{(i)} \}$ (referred to as configurations)
which sample the probability
distribution ${\vert \Psi_T({\bf R}) \vert^2
/ { {\cal N}}}$ and where $\hat{H}\Psi_T({\bf R}_i)
/ {\Psi_T({\bf R}_i)}$ is referred to as the local energy, which
is summed over a suitably large number of configurations, $n$.
These configurations are generated using the 
Metropolis algorithm.\cite{Metroplis}

By the variational principle we know that this expectation value is an
upper bound on the exact ground-state energy.
To get as close as possible to the ground-state,
we minimize
$\sigma_{l.e.}$,\cite{Umrigar,Coldwell} the standard deviation of the
local energy.

The trial wavefunction, $\Psi_T$ is written as a product of spin-up
and spin-down Slater
determinants, $D^\uparrow$ and $D^\downarrow$,
containing LSDA orbitals calculated for the same system.
In addition the wavefunction contains a Jastrow factor
\cite{Jastrow} which describes
the correlations between the electrons. For our Jastrow factor we use
the same as that used by Sottile {\it et al}.
\cite{sottile} in their calculations of jellium spheres
(they did not include an embedded atom), except for
a 'multipolar' term which we do not include in our calculations.
The trial wavefunction is

\[
\Psi_T({\bf r}_1,\cdots,{\bf r}_N)=\exp \left[ \sum_{i=1}^N\left(\sum_{n=1}
^6\alpha_n^{(i)}j_0\left({n\pi r_i \over{R_{jell}}}\right) \right)\right]
\]
\[
\times \exp\left[\sum_{1\leq i < j \leq N} \frac{1}{2}\left( 
{{a_{ij}r_{ij}+b_{ij}{r_{ij}}^2}\over{1+c_{ij}(r_i)r_{ij}+
d_{ij}{r_{ij}}^2}}+\right.\right.\]
\[
\left.\left.
{{a_{ij}r_{ij}+b_{ij}{r_{ij}}^2}\over{1+c_{ij}(r_j)r_{ij}+
d_{ij}{r_{ij}}^2}}\right) \right]
\]
\[ 
\times D^\uparrow({\bf r}_1,\cdots,{\bf r}_{N/2})
D^\downarrow({\bf r}_{N/2+1},\cdots,{\bf r}_{N})
\] 

\noindent where $c_{ij}$ is given by

\[ 
c_{ij}(r_i)=c_{ij}^0+c_{ij}^1\arctan[(r_i^2-R_{jell}^2)/2
\bigtriangleup R_{jell}]
\label{eq:c_definition}
\] 

Parameters $a_{ij}$, $b_{ij}$, $c_{ij}^0$, $c_{ij}^1$ and $d_{ij}$ 
depend only on the relative spins of electrons $i$ and $j$
and $\alpha_n^{(i)}$ depend only on the spin of electron $i$. 
The parameters $a_{ij}$ are
determined by the electron-electron cusp condition. 
Note that the nuclear-electron cusp condition
is automatically satisfied on account of the LSDA
orbitals in the determinant.
The other 15 
parameters in the Jastrow factor are variational parameters
which are varied in order to minimize
$\sigma_{l.e.}$. Correlated sampling
\cite{var_dif_review,corr_samp,corr_samp2} is used in the
minimization procedure, which avoids the need to
recalculate new configurations for each set of Jastrow parameters.

\subsection{Fixing the jellium sphere size}

The number of electrons in the jellium sphere is chosen so that the
immersion energy is as close to that of the atom in infinite jellium
as possible. In addition, in order to have a good trial wavefunction
we require a number of electrons
such that it is
possible to choose a configuration of the electrons in the open-shell
so as to give a purely real wavefunction overall.
The VQMC energy is also minimized by choosing configurations which are
as non-magnetic as possible. With these considerations 
in mind, and also with a view to minimize computational effort, we choose
a 10-electron jellium sphere for the calculations presented below.
This choice leads to the following VQMC electron configuration for the
Hydrogen in jellium

\[
1s^{1\uparrow}, 1s^{1\downarrow}, 2s^{1\uparrow}, 2s^{1\downarrow},
2p^{3\uparrow}, 2p^{3\downarrow}, 3d^{1\uparrow}
\]

\noindent and the configuration for the 10-electron
jellium sphere without the Hydrogen atom

\[
1s^{1\uparrow}, 1s^{1\downarrow}, 
2p^{3\uparrow}, 2p^{3\downarrow}, 3d^{\uparrow}, 3d^{\downarrow}
\]

\noindent where the $d$ electrons are placed in orbitals with magnetic
quantum number, $m=0$.

In the LSDA solution for the 10-electron jellium sphere, the two $3d$
electrons are instead filled as $3d^{2\uparrow}$ in order to minimize the
total energy, in a manner consistent with Hund's rule.
Note that the $d$ sub-shell is not full
and so the density contribution from these
electrons is spherically symmetrized in LSDA before recalculating
the KS potentials.

\section{\label{sec:SIC}SIC Solution}

In our SIC calculations, the SIC will not be applied to all electrons in
the system. Instead
we apply SIC only to the $1s$ spin-up and spin-down electrons.
This approach is taken because our Hydrogen 
atom in a jellium sphere is meant to
be an approximation of Hydrogen in infinite jellium, and in the latter, only
the $1s^\uparrow$ and $1s^\downarrow$ electrons are sufficiently localized
to warrant the correction.
Hence for our SIC solution, the $1s$ electrons will obey KS equations 
with a SI-corrected potential and
all the other electrons obey the standard LSDA KS equations.

For our 10-electron jellium sphere,
the $1s$ and $2s$ orbitals of a given spin are calculated
from different eigenvalue equations (unlike in the LSDA, where both
share the same potential) and so will not in general be orthogonal.
However, the KS formulation of DFT requires this orthogonality and so it
must now be imposed, for which we use Gram-Schmidt orthogonalization
\cite{Schmidt}

\[ 
\psi_{2s}^{\sigma, orth} (r)={\cal{N}}^\sigma \left(\psi_{2s}^{\sigma} (r)-
\int \psi_{1s}^{\sigma *} (r')
\psi_{2s}^{\sigma} (r') d{\bf r}' \psi_{1s}^{\sigma} (r) \right)
\] 

\noindent where $\cal{N}^\sigma$ ensures the correct normalization
of the orbitals.

The SIC energy functional is

\[
E=\sum_{n \ne 2,l,m,\sigma}\int  
{\psi_{nlm}^{\sigma *}}
({\bf r})(-\frac{1}{2} \nabla^2) {\psi_{nlm}^{\sigma}} ({\bf r})d{\bf r}
+\]
\[
\sum_\sigma \int  
{\psi_{2s}^{\sigma, orth *}}
({\bf r})(-\frac{1}{2} \nabla^2) {\psi_{2s}^{\sigma, orth}} ({\bf r})d{\bf r}+
\]
\[
\sum_{m,\sigma}  \int  
{\psi_{2p,m}^{\sigma *}}
({\bf r})(-\frac{1}{2} \nabla^2) {\psi_{2p,m}^{\sigma}} ({\bf r})d{\bf r}+
\]
\[
\frac{1}{2} \int \int  
{n(r)n(r')
 \over {\vert {\bf r} - {\bf r}' \vert}} d{\bf r} d{\bf r}'+
\int v_{ext}(r)
n(r) d{\bf r}+E_{xc}[n^{\uparrow},n^{\downarrow}]-\]
\[
\sum_{\sigma}\left[ \frac{1}{2} \int \int  
{n_{1s}^\sigma(r)n_{1s}^\sigma(r')
 \over {\vert {\bf r} - {\bf r}' \vert}} d{\bf r} d{\bf r}'
+E_{xc}[n_{1s}^\sigma,0]
\right]
+
\]
\[ 
Z \int_{r=0}^{r=R_{jell}} { n_0 \over { r }} d{\bf r}
+\frac{3}{5} {N^{5/3} \over {r_s}}
\] 

Note that the $2s$ contribution to the kinetic energy is most easily
calculated by evaluating the radial derivatives of the orbitals. The
alternative method of substituting in from the KS equations is also possible,
but is more complicated in
this case as the orthogonalized orbitals are now mixtures of the eigenstates
of the KS equations. 
As with the LSDA results, both methods are found to give the same results.

\section{\label{sec:results}Results}

\subsection{\label{sec:Total energies and immersion energies}
Total energies and immersion energies}

\begin{figure}
\scalebox{0.8}{{\includegraphics{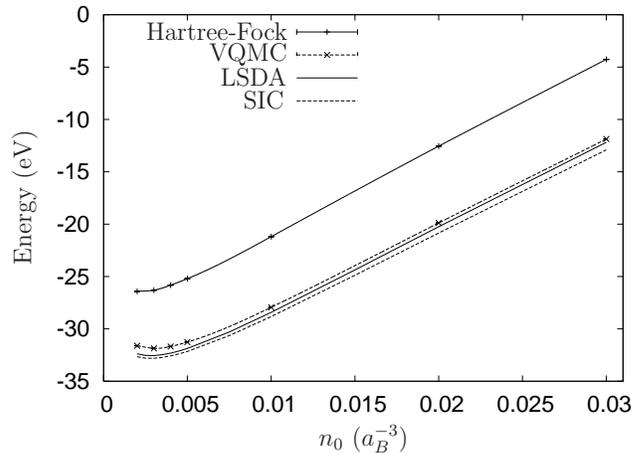}}}
\caption{\label{fig:energies_ainj_10}
Total energy of a Hydrogen atom immersed in a 10-electron
jellium sphere for different background densities.
}
\end{figure}

We report VQMC solutions for a Hydrogen atom immersed in
10-electron jellium spheres 
of background densities $0.001a_B^{-3}$ through to $0.03a_B^{-3}$
($r_s$ values of $6.2$ and $2$ respectively).
Total energies were calculated and are presented, along with LSDA
and SIC results, in Fig.~\ref{fig:energies_ainj_10}.
Total energies are also calculated using a close approximation to
the Hartree-Fock method (HF). These points are calculated by evaluating
the expectation value of the Hamiltonian for a wavefunction 
consisting of just the Slater determinant part containing
the LSDA orbitals. This is achieved by performing a VQMC run with the Jastrow
factor set equal to one.

\begin{figure}
 \scalebox{0.8}{{\includegraphics{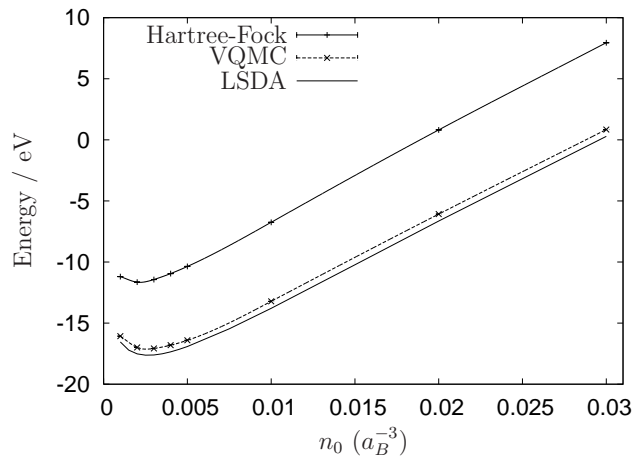}}}
 \caption{\label{fig:energies_jell_10}
 Total energy of a 10-electron
 jellium sphere for different background densities.
 }
 \end{figure}

\begin{figure}
 \scalebox{0.8}{{{\includegraphics{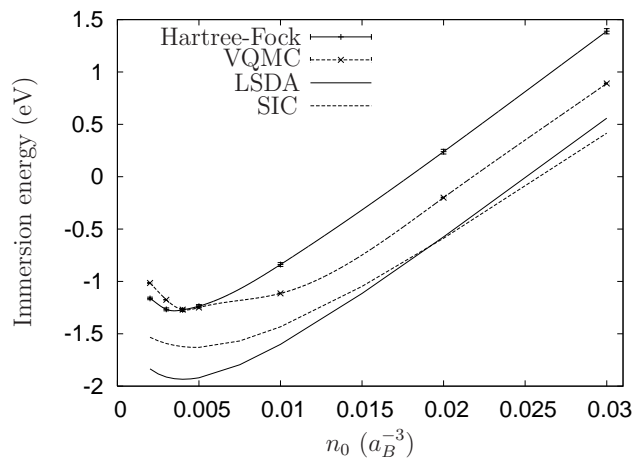}}}}
 \caption{\label{fig:imm_energies_H_10}
 Immersion energies for a Hydrogen atom immersed in a 10-electron
 jellium sphere for different background densities.
 }
 \end{figure}

Energies for jellium spheres but without the Hydrogen atom
are presented in Fig.~\ref{fig:energies_jell_10}. 
Finally, the immersion energies
are reported in Fig.~\ref{fig:imm_energies_H_10}.

In calculating the immersion energy for HF, VQMC and SIC
the exact value of the Hydrogen atom energy is used, namely $-13.606eV$. This is appropriate as calculations using HF and 
SIC both give this value and in the case of VQMC, 
the calculation is correct to a very high level of precision.
For the LSDA calculation, the atom energy is found to be $-13.030$eV.

The total energy curves calculated using the different methods are 
all of the same shape and feature minima at roughly the same
locations. The LSDA, SIC and VQMC curves lie within $1eV$
of one another in the atom in jellium total energy
graph and similarly the LSDA and VQMC curves 
are within $1eV$ of one another in the jellium sphere total energy graph.
HF gives total energy curves for the atom in jellium
and for the jellium sphere which are rigidly shifted by about $5$eV above the
VQMC curves.

We note that our VQMC total energy curve for the jellium sphere is
slightly different to the type obtained by Sottile {\it et al}.\cite{sottile}
Their VQMC results were found to be
slightly lower than the LSDA results above background
densities of about $0.01a_B^{-3}$.
This is because
we neglected to include the 'multipolar' term in the trial wavefunction
which Sottile {\it et al}. \cite{sottile} used in their calculations. In their work, 
this term was found
to improve the VQMC solution for the larger background densities, but
was relatively unimportant at densities near the minimum.

When the total energy curves are subtracted from one another 
to give the immersion
energy curves, the large error in the HF calculations cancel out.
We find all four curves lie within $1$eV of one another and
all curves feature a minimum at approximately $0.004 a_B^{-3}$.
Notice that the LSDA results for the immersion energy curve 
are lower than the VQMC results, which means that the LSDA
is overbinding the atom to the jellium relative to
VQMC.

Calculations by Sugiyama {\it et al}. \cite{Sugiyama} 
of the immersion energy has a discrepancy
of 5eV between Monte Carlo and LSDA, which is very much larger than we find
here. Possibly this difference was because of a different choice of trial
wavefunction.

We see that the VQMC immersion energy curve
below $0.005 a_B^{-3}$ is more closely followed by the
SIC curve than by the LSDA curve. 
If we regard our VQMC results as a benchmark (see Section \ref{sec:conc}
for a discussion of this) then this indicates that the SIC is
outperforming the LSDA for these background densities.
This is expected as the
$1s$ bound states are becoming more localized at these low
background densities and therefore
the self-interaction of these bound states is increasing.

Note that for an atom in infinite jellium the SIC and LSDA curves
would begin to coincide at large background densities
due to the $1s$ electrons becoming increasingly delocalized.
However for our system of an atom in finite jellium this is not the case.
As the background density is increased, the sphere size becomes smaller
and so effects due to the finite size of our system begin to impinge on the
results.

\subsection{\label{sec:Atom-induced densities}
Atom-induced densities}

\begin{figure}
\scalebox{0.8}{{{\includegraphics{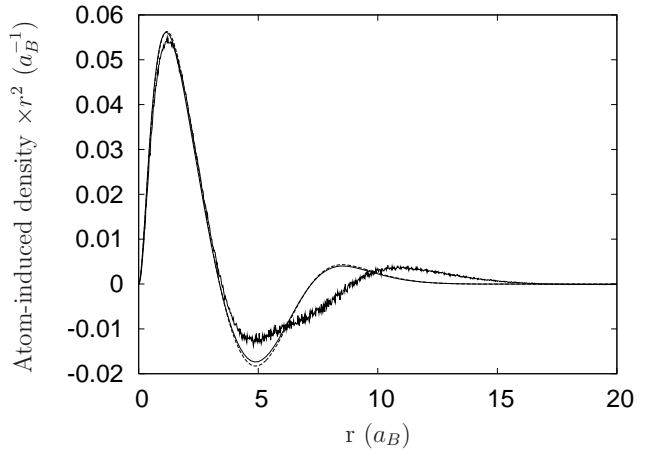}}}}
\caption{\label{fig:atom induced densities}
Atom-induced density as calculated using the different
methods, for $n_0=0.004a_B^{-3}$. The solid line is the LSDA density,
the dashed line is the SIC density and the noisy line is the VQMC density.}
\end{figure}

Atom-induced densities (multiplied by $r^2$)
are plotted in Fig.~\ref{fig:atom induced densities} for a background
density of 0.004$a_B^{-3}$.
The Hydrogen-like maximum in the atom-induced density is very similar
between VQMC and LSDA for $r$ less than $4a_B$.
A discrepancy of at most 5\% is identified in this region.
On the other hand, as highlighted by Hoffman and Pratt,\cite{Hoffman}
the work of Sugiyama {\it et al}. \cite{Sugiyama} shows
a difference of 30\% between LDA and VQMC
atom-induced densities near the proton.
The much smaller discrepancy in our results
is probably due to our use of a
trial wavefunction which performs better than that of
Sugiyama {\it et al}. close to the proton.

However, our atom-induced density plots show the same discrepancy 
in the Friedel oscillation between LSDA and
VQMC as was seen by Sugiyama {\it et al}.\cite{Sugiyama,Hoffman} 
In particular the minimum in the atom-induced density for the
LSDA is much deeper than that for the VQMC.
Also the wavelength of the
oscillation is larger for VQMC and so
the Friedel oscillation maximum occurs at a larger radius.

The SIC atom-induced density is also included in 
Fig.~\ref{fig:atom induced densities}. The SIC and LSDA densities
are essentially identical, which seems unusual given the difference in the 
immersion energies for these methods.

\section{\label{sec:conc}Conclusions}

We have presented what we believe to be the first
reliable VQMC results for the atom in jellium model system.
The immersion energy versus background density curves for VQMC,
LSDA and SIC lie within $1eV$ of each other with minimas in approximately
the same positions. 
The LSDA results show an overbinding of the atom to the jellium relative to
the VQMC result. Viewing the VQMC as a benchmark, this is consistent
with the general overbinding seen in LSDA.
In addition, for low background densities,
the immersion energy curve of VQMC is more closely followed by the
the SIC immersion energy curve than by the LSDA immersion energy curve.
Again, viewing VQMC as a benchmark,
this is consistent with the fact that the SIC is expected to be more
accurate than the LSDA for systems with more strongly localized electrons
(as is the case for the low background densities).

The status of the VQMC results as a benchmark is uncertain however.
More accurate DQMC calculations would yield lower total
energies of both the atom in jellium and the jellium
relative to the VQMC results. However,
because the immersion energy is calculated as the difference 
between these energies, part of the change in going from VQMC to DQMC will
cancel out when we calculate the immersion energy.
Furthermore,
Sottile {\it et al}. \cite{sottile} found a rigid shift of only $-0.2$eV in the DQMC
energies of an 8-electron jellium sphere relative to the VQMC energies.
This suggests that our results for the immersion energy would not be
strongly modified by using DQMC.

\section*{ACKNOWLEDGEMENTS}

We would like to acknowledge EPSRC and STFC funding and 
also helpful discussions
with B. Gy\"orffy, W. Temmerman, Z. Szotek and M. L\"uders.

\bibliography{text}

\end{document}